\documentclass[%
preprint,%
]{revtex4-1}

\usepackage{graphicx}
\usepackage{dcolumn}
\usepackage{bm}
\usepackage{amsfonts}
\usepackage{bbold}
\usepackage[utf8]{inputenc}
\usepackage[T1]{fontenc}
\usepackage{mathptmx}
\usepackage{subfigure}
\usepackage{epigraph}
\usepackage{float}
\usepackage[top=2cm, left=2.5cm, right=2cm, bottom=2cm]{geometry}
\usepackage{amsmath, latexsym}
\usepackage{mathtools}
\usepackage{amssymb}
\usepackage{xparse}
\usepackage{physics}
\usepackage{color}
\usepackage{stackrel}
\usepackage{tikz}
\usetikzlibrary{arrows}
\usepackage{wrapfig}
\usepackage[english]{babel}

\newcommand{\midarrow}{\tikz \draw[-triangle 90] (0,0) -- +(.1,0);}

\begin{document}

\preprint{SB/F/480-18}

\title{The $SU(3)$ $A_1$ graph and its associated quantum groupoid.}

\author{M. Dias}
\email{mida.dias01@gmail.com}
\author{E. Isasi}%
 \email{Corresponding Author: eisasi@usb.ve }

\author{L. Vásquez}
\email{luis.fourth@gmail.com}
\affiliation{Physics Department, Simón Bolívar University.}
\date{\today}

\begin{abstract}
An explicit and complete construction of the $SU(3)$ $A_{1}$ associated  quantum groupoid is presented in this work, inspired by the approach taken by Trinchero\cite{Trinchero:2010yr} for the $SU(2)$ $A_l$ graphs. New creation and annihilation operators were defined in order to consider the $3$ different types of back-tracks which appear due to the specific structure of $SU(3)$. The $C^{\star}$ bialgebra and the realization of a Temperley-Lieb algebra is studied thoroughly. Finally, it is shown that the construction of the quantum groupoids associated to the $A_{1}$ $SU(N)$ graphs are easily obtained for any value of $N$ using the results of this work. The generalization for higher levels $A_l$ graphs are still an unsolved challenge, but now we count with enough tools, some insight about how to attack this problem, and the first steps towards solving it.
\end{abstract}

\maketitle

%

\section{\label{sec:level1}Introduction}

The existence and explicit construction of the quantum groupoids associated to a $SU(N)$ $ADE$ graphs, is an interesting and complex mathematical problem that has been addressed for several years, using many different points of view \cite{Ocneanu:paths,CoqueTrincheroDTE:2004,CoqueA2:2005, IsasiSchieber:2007, HammahuiA2SU32008, Evans:2009ud,Trinchero:2010yr,CoquereauxIsasiSchieber:2010,CaicedoIsasiPineda:2015,kuperberg1996}. 
The applications and different connections of this structures occupy an important part of the Mathematical Physics literature, specially in Integrable Systems and Conformal Field Theory \cite{JonesGoodmanTowers:1989,DiFrancescoSenechal:1997}. 

However, when we look  at the $SU(3)$ family of graphs there is only one attempt \cite{HammahuiA2SU32008} to explicitly construct the entire structure. In that work, the quantum groupoid of the $A_1$ graph is studied using the language of paths over the graph and Ocneanu Cells. Despite the fact that all the structure seems to be present in Hammaoui's work \cite{HammahuiA2SU32008}, when one take a closer look, one realizes that the algebra is not complete, basically due to the fact that there are paths which are not taken into account. In the language we use in this work and in Hammaoui's work \cite{HammahuiA2SU32008} this space is the free vector space spanned by the set of graduated endomorphisms of paths on the $A_1$ graph. The consequence is crucial because the product so defined, even thought it is closed and consistent (in the sense of homomorphism) with the co-product and the remaining bi-algebra operations, is non-associative. 

For higher level and exceptional graphs the only known method to recover the information of these quantum groupoids uses the Modular Splitting Method \cite{IsasiSchieber:2007} which allows to recover the corresponding matricial representation.

For the $SU(2)$ case a general, very elegant and explicit definition of the operations of the associated quantum groupoids has been constructed by Trinchero \cite{Trinchero:2010yr}.  Using the path language in Trinchero's work \cite{Trinchero:2010yr} it is shown that for any simple $ADE$ bioriented graph it is possible to construct the quantum groupoids directly from the properties of the space of paths without having to use Ocneanu cells. The key ingredient is the decomposition of the space of paths as a direct sum of subspaces which are: either the subspace of essential paths with $n$ steps or orthogonal subspaces constructed by recurrent applications of the corresponding creation operators $C_i^{\dagger}$ ($U_i=C_i^{\dagger}C_i$) on subspaces of essential paths of shorter length. This decomposition and the corresponding orthogonal projectors, are sufficient to define all the operations of the quantum groupoids for any $ADE$ graph and any afine $ADE$ graph.

In this work we present a complete construction of the quantum groupoid defined over the free vector space of graded endomorphisms of paths over the simply laced graph $A_1$ of $SU(3)$. The construction imitates Trinchero's development \cite{Trinchero:2010yr}, introducing several important differences specific of $SU(3)$ that we think are critical for the understanding of the higher levels. The important differences with $SU(2)$ appear because the $SU(3)$ bialgebra is constructed over two oriented graphs, one for each of the two $SU(3)$ fundamental generators. This implies that $SU(3)$ requires three types of operators acting on 3 different types of back-tracks or spurious sections of a path, instead of just one operator as in $SU(2)$. One of these new operators considers triangular sections of two contiguous edges, generated by the repeated action of two generators $\sigma$ ($\bar{\sigma}$ in the conjugate graph), which will be replaced by an edge generated by $\bar{\sigma}$ ($\sigma$ in the conjugate graph)\footnote{For a geometric description of this back-track and the action of the operators see Pineda \cite{CaicedoIsasiPineda:2015} et al. and Conquereaux \cite{CoquereauxIsasiSchieber:2010}} et al., and the other two operators that will act on back-tracks of type $SU(2)$ composed of an edge in each direction generated by the action of the generators in the form $\sigma\bar{\sigma}$ (or $\bar{\sigma}\sigma $).

As far as we know, at the moment this is the only explicit case of the complete construction of the weak Hopf algebra (associated to $SU(3)$) available in the literature. Moreover, if we look at the $A_{1}$ graphs of $SU(N)$ for some $N$, one finds that the process to construct the corresponding quantum groupoids is exactly the same as for $SU(3)$ but requiring more operators, given that there are more generators and therefore, more types of back-tracks.  Geometrically, this is actually presented in such a way that it only remains to prove a reduced (but complex) number of properties to obtain the $C^\star$ bialgebra for any level, including exceptional cases.

The paper is organized as follows. In  section \ref{section2}, the $SU(3)$ $A_{1}$ associated graph is thoroughly described. Then, in section \ref{section3}, the corresponding Jones operators $U_{i}$ are defined from the construction of the corresponding creation and annihilation operators, and it is proven that these are the elements of a Temperley-Lieb Algebra. Next, in section \ref{section 4}, a $C^{\star}$ bialgebra is constructed by using the graded endomorphisms over the space of paths of $A_{1}$ and in section \ref{grafos A1} we generalize this construction to the $A_{1}$ graph of any $SU(N).$ Finally, in section \ref{section: higherlevelA}, some ideas about the higher levels $A_{l}$ graphs of $SU(3)$ are discussed.

\section{The $SU(3)$ $A_1$ graph}
\label{section2}

In general, the family of $SU(N)$ $A_l$ graphs are the Weyl alcoves truncated at some level $l$, and are characterized by its Coxeter number $\kappa=N+l$. The vertices of the graphs denote irreducible representations (irreps) of the quantum sub-groups $SU(N)_l$ at roots of unity $q=e^{i\frac{\pi}{\kappa}}$ 
\cite{DiFrancescoSenechal:1997}, \cite{Fuchs:1992}, \cite{Ocneanu:MSRI}. 

The Weyl alcoves of $SU(3)$ are two dimensional, simply laced, oriented graphs, with a structure of triangular mesh. The vertices (the irreducible representations of $SU(3)_l$) can be labelled using triangular coordinates
{\small
\begin{equation}
\{\lambda=(\lambda_1,\lambda_2)=\lambda_1\Lambda_1+\lambda_2\Lambda_2\, s.t.\, 0\leq \lambda_1+\lambda_2\leq l,\, \lambda_{1,2}\in\mathbb{N}\},
\end{equation}}

where $\Lambda_1$ and $\Lambda_2$ are the fundamental weights of the $SU(3)$ Lie group, $\lambda_1$ and $\lambda_2$ are the corresponding Dynkin labels, and the rule $0\leq \lambda_1+\lambda_2\leq l$ gives the appropriate truncation at level $l$ to obtain irreducible representations of $SU(3)_l$. With this labelling the vertex $(0,0)$ is the unit representation related to the ''vacuum state'', $(1,0)$ is the fundamental representation (generator) of $SU(3)_l$ and $(0,1)$ its conjugate representation (also a generator)\cite{CoquereauxHammahui:2006}. This means that $SU(3)_{l}$ is associated with two oriented graphs (which are isomorphic but with opposite orientation), one generated by the fundamental representation $(1,0)$ and one generated by the conjugate fundamental representation. The cardinality of $A_l$ is $d_{A_l}=\frac{(l+1)(l+2)}{2}$. 

The quantum dimension of a given vertex $\lambda = (\lambda_1 ,\lambda_2)$ is given by the q-analog of the classical formula for dimensions of irreps of $SU(3)$, real numbers being replaced by quantum numbers:\cite{CoquereauxHammahui:2006} 
\begin{equation}qdim(\lambda) = (1/[2]_q )([\lambda_1+ 1]_q [\lambda_2+ 1]_q [\lambda_1+\lambda_2+ 2]_q),\end{equation}
where $q = exp(i\pi/l)$ is a root of unity and $[n]_q = \frac{q^n-q{-n}}{q-q^{-1}}$\cite{CoquereauxHammahui:2006}. The quantum dimension $[2]_q$ is special because appears as the deformation parameter in the Temperley- Lieb algebra, and it is denoted by $\beta$.

With this definition the vertices of the graphs can be also labelled using the quantum dimensions. Grouped by levels, the first dimensions will be $\{\{[1]_q\},\{[3]_q,[3]_q\},\{[6]_q,[8]_q,[6]_q\}\dots\}$ that we will denote with the short notation $\{\{1\},\{3,\bar{3}\},\{6,8,\bar{6}\}\dots\}$ where the bar was added to identify the conjugate representation.

For the specific case of $SU(3)$ $A_1$, the graphs have two levels with quantum dimensions $\{\{[1]_q\},\{[3]_q,[3]_q\}\}$ and the vertices of the graphs are denoted using $\{\{1\},\{3,\bar{3}\}\}$. Notice that for $l=1$ we have  $[1]_q=[3]_q=1$ and then the actual dimension appearing on the vertices is the real number $1$. The graphs is $Z_3$ symmetric under rotations of $120^{\circ}$ as expected. 

In order to make our discussion clearer and more precise we will differentiate the two associated oriented graphs. We denote $\overrightarrow{A}_1$ as the graph generated by the fundamental representation $(1,0)$ and $\overleftarrow{A}_1$ as the graph generated by the conjugate fundamental representation $(0,1)$. The graph $A_{1}$ will have both the edges of $\overrightarrow{A_{1}}$ and $\overleftarrow{A_{1}}.$

\section{Realizing the $SU(3)$ Temperley-Lieb algebra in terms of paths}
\label{section3}

Considering the graphs $\overrightarrow{A}_1$ and $\overleftarrow{A}_1$, an elementary path will be a sequence of vertices $(v_{0},..., v_{n})$ such that each pair of consecutive vertices have an edge connecting them either in $\overrightarrow{A_1}$ or in $\overleftarrow{A_1}$. 

\begin{figure}[H]
\centering
\begin{tikzpicture}
\begin{scope}[very thick, every node/.style={sloped,allow upside down}]
  \draw (0,0) node[below] {1}-- node {\midarrow} (2,0);
  \draw (2,0) node[below] {3}-- node {\midarrow} (1,1.7) ;
  \draw (1,1.7) node[above] {$\overline{3}$} -- node {\midarrow} (0,0);
\end{scope}
\end{tikzpicture}
\caption{ \label{fig:A1} The $A_1$ $SU(3)$ graph generated by the fundamental representation $(1,0)$ with the vertices labeled by the short notation for quantum dimensions.}
\end{figure}
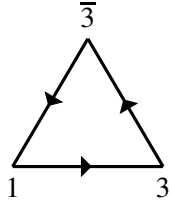

The number $n$ will be called the number of steps in the path. Let's define an inner product such that each pair of different elementary paths are orthogonal and then, define a free-vector space $\mathcal{P}$ with that inner product, where an arbitrary path is a formal linear combination of elementary paths. It is natural to decompose this space of paths into orthogonal subspaces $\mathcal{P}_{n}$ which contain all the paths with $n$ steps. From now on, an edge which belongs to $\overrightarrow{A_1}$ will be called a ``forward step'', while an edge which belongs to $\overleftarrow{A_1}$ will be called a ``backward step''. The length of a path will be the ordered pair $(x,y)$ such that $x$ is the number of forward steps in the path and $y$ is the number of backward steps. Furthermore, we can define orthogonal subspaces $\mathcal{P}_{i,j}$ containing all the paths of length $(i,j)$.  

Here we present how to define the Jones operators $U_{i}:\mathcal{P}_{n}\rightarrow \mathcal{P}_{n}$ (the $i$ stands for the $ith$ step of the path in which it is applied), such that these constitute a realization of the $SU(3)$ Temperley-Lieb algebra. This is equivalent to ask that the $U_{i}'s$ satisfy

\begin{gather}
    U_{i}^{2}=\beta U_{i},
\label{TL 1} \\
    U_{i}U_{i+1}U_{i}-U^{i}=U_{i+1}U_{i}U_{i+1}-U_{i+1},
\label{TL 2}\\
    U_{i}U_{j}=U_{j}U_{i}\quad \text{for } |i-j|>1,
\label{TL 3}\\
    (U_{i}-U_{i+2}U_{i+1}U_{i}+U_{i+1}) (U_{i+1}U_{i+2}U_{i+1}-U_{i+1})=0
    \label{TL 4}
\end{gather}

Notice that, in these expressions, we must consider the number of steps of the path on which we are acting. For example, if we are verifying these expressions for $3$-step paths, we should only consider operators $U_{0}$ and $U_{1}$.

\subsection{Creation and annihilation operators in $SU(3)$}
In the $A_1$ graph, we have three elementary paths of length $(0,0)$, three elementary paths of length $(1,0)$, and three elementary paths of length $(0,1)$. We define the sub space of essential paths $\mathcal{E}$ bay saying that the previous set of paths constitute an orthonormal basis of $\mathcal{E}$ with the same inner product as $\mathcal{P}$. The reason behind this is that there are no shorter ways to join any two vertices of the graphs under consideration. The shortest way to join any vertex with itself is a path with zero steps. The shortest way to join two different vertices is always an edge, which is either a path of length $(1,0)$ or $(0,1)$. Similarly, $\mathcal{E}_{n}$ will be the subspace of essential paths with $n$ steps. Also, we will denote the subspace of essential paths of a given length $(i,j)$ as $\mathcal{E}_{i,j}$. We will see in what follows that this definitions coincides with the original definition of essential path, i.e. the set of paths that are annihilated by all the Jones operators, provided that a good definitions of these last ones is available. In this sense this definition can also be read as an ansatz.

We define creation operators such that every possible path with $2$ steps can be created from the essential paths. One way to do this is the following. Let's define $3$ types of creation operators for $n\geq 0$ 
\begin{gather}
T_{i}^{\dagger}:\mathcal{P}_{n+1}\rightarrow \mathcal{P}_{n+2},\\
B_{i}^{\dagger}:\mathcal{P}_{n}\rightarrow \mathcal{P}_{n+2}, \\
F_{i}^{\dagger}:\mathcal{P}_{n}\rightarrow \mathcal{P}_{n+2},
\end{gather} 
where it can be seen that $T^{\dagger}$ cannot be applied over a path with zero steps. 

Over a path $\eta=(v_{0},v_{1},...,v_{n})$ with the right number of steps, each of these operators will act in the following ways:
\begin{gather}
T^{\dagger}_{i}(\eta)=\sum_{j}\overline{t}_{v_{i}j v_{i+1}}(v_{0}...v_{i},j,v_{i+1}...v_{n}),
\label{t+ definicion} \\
F^{\dagger}_{i}(\eta)=\sum_{j}\overline{f}_{v_{i}j v_{i}}(v_{0}...v_{i},j,v_{i},v_{i+1},...v_{n}),
\label{f+ definicion}\\
B^{\dagger}_{i}(\eta)=\sum_{j}\overline{b}_{v_{i}j v_{i}}(v_{0}...v_{i},j,v_{i},v_{i+1},... v_{n}),
\label{b+ definicion}
\end{gather}
where $\overline{t}_{ijk},$  $\overline{f}_{ijk},$ and $\overline{b}_{ijk}$ are complex numbers. These constants are fixed in a way in which the $T^{\dagger}$'s add triangle-like paths of length $(0,2)$ or $(2,0)$, while the rest of the operators add loops of length $(1,1)$. The $F^{\dagger}$'s operators add loops which start with a forward step, and the $B^{\dagger}$'s operators add loops which start with a backward step. Using the rule:
\begin{equation} (C^{\dagger}_{i,j}\eta, \eta')=\overline{(\eta,C_{i,j}\eta')}, \end{equation}
we can find the associated annihilation operators. Notice that we use $C_{i,j}$ to denote the operators previously defined. The $i$ stands for the type of operator, that is $i=0,1,2$ corresponds to $T$, $F$ and $B$ respectively. Similarly, the $j$ stands for the step in which the operator is applied. The annihilation operators are then:
\begin{gather}
T_{i}(\eta)=t_{v_{i}v_{i+1}v_{i+2}}(v_{0},v_{1},...,v_{i},v_{i+2},..v_{n}),
\label{t definicion}\\
F_{i}(\eta)=f_{v_{i}v_{i+1}v_{i+2}}(v_{0},v_{1},...,v_{i},v_{i+3},..v_{n}),
\label{f definicion}\\
B_{i}(\eta)=b_{v_{i}v_{i+1}v_{i+2}}(v_{0},v_{1},...,v_{i},v_{i+3},..v_{n}).
\label{b definicion}
\end{gather}
The $T$'s replace triangle-like parts of the path for an edge. In the same way, the rest of the operators erase loops of length $(1,1)$.

These operators have some similarities with the creation and annihilation operators of quantum mechanics. In fact, the elementary paths behave as number states and the essential paths as the corresponding vacuum states.  

\subsection{The Temperley-Lieb algebra}

Now, inspired by the resemblance with quantum mechanics, we can define the Jones operators $U_{i}:\mathcal{P}_{n}\rightarrow \mathcal{P}_{n}$ which will have a similar form to the number operators. Let's define:
\begin{equation}
U_{i}=T^{\dagger}_{i}T_{i}+B^{\dagger}_{i}B_{i}+F^{\dagger}_{i}F_{i}.
\label{e definicion}
\end{equation}
If we require that these operators lead us to a Temperley-Lieb algebra, then, we immediately notice that
\begin{equation}\beta=|t_{v_{2}v_{1}v_{0}}|^{2}=|t_{v_{0}v_{1}v_{2}}|^{2}=|f_{v_{0}v_{1}v_{0}}|^{2}=|b_{v_{1}v_{0}v_{1}}|^{2},
\label{constantes}
\end{equation}
for any path $\eta=(v_{0},v_{1},v_{2})$ of length $(2,0)$. We can see this by evaluating equation (\ref{TL 1}) in elementary paths of length $(1,1)$, $(2,0)$, and $(0,2)$. This implies for this particular case that

\begin{equation}U_{i}(\eta)=\beta \eta \qquad \text{for } i<n-2.\end{equation}
Therefore, the equations (\ref{TL 2}) and (\ref{TL 3}) are automatically satisfied. Equation (\ref{TL 4}) can only be satisfied if $\beta$ is $0$, $1$, or $\sqrt[]{2}$. Notice that, if we choose $\beta=\sqrt[]{2}$, then equation (\ref{constantes}) is in agreement with the computation of triangular cells presented in Conquereaux \cite{CoquereauxIsasiSchieber:2010} et al. and Evans \cite{Evans:2009ar} et al. This also happens in the $A_1$ graph associated with $SU(2)$, which is two vertices joined by an edge. The only possible paths include sequences of 2-length loops, and nothing more. Therefore, the number operator is proportional to the identity when acting on paths of length greater than or equal to $2$.

\section{The Bialgebra}
\label{section 4}

Now, knowing that we can realize a Temperley-Lieb algebra structure for $A_1$ of SU(3), we want to construct the associated weak $\star$ Bialgebra, inspired by the approach taken in Trinchero's work \cite{Trinchero:2010yr}, modified in a way that it takes into account the $3$ different kind of back-tracks which we have in $A_1$ of $SU(3)$. The proofs are made in such a way that if we manage to define correctly the previously mentioned operators for any $A_l$ graph, together with the corresponding decomposition of the space of paths, the construction of the weak $\star$ Bialgebra will be similar for higher levels. \footnote{For the next levels, we are convinced that we can discover the right operators to make the decomposition by studying the $A_1$ graph of a particular $SU(N)$. See section \ref{section: higherlevelA} for a more detailed description.}

We must first define the vector space of graded endomorphisms of paths over $A_1$ as
 \begin{equation}
 End^{gr}(\mathcal{P})=\stackbin[i+j=0]{\infty}{\bigoplus}(\mathcal{P}_{i,j}\times \mathcal{P}_{i,j}).
\end{equation}

 Similarly, we must also define the vector space of graded endomorphisms of essential paths over $A_1$ as
\begin{equation}
\mathcal{A}\equiv End^{gr}(\mathcal{E})=\stackbin[i+j=0]{l}{\bigoplus}(\mathcal{E}_{i,j}\times \mathcal{E}_{i,j}),
\end{equation}
where $l$ is the maximum number of steps of an essential path, which is actually $1$ for $A_1$. These graded endomorphisms of essential paths constitute a weak $\star$ bialgebra if we equip them with the right operations.  

\vskip 0.3cm

To start, let $\star$ denote the concatenation product of paths. If $\eta=(v_{0},...,v_{n})$ and $\eta'=(\omega_{0},....,\omega_{m})$, then
\begin{equation}
    \eta \star \eta ' =\delta_{v_{n}\omega_{0}}(v_{0},...,v_{n},\omega_{1},...., \omega_{m}).
\end{equation}

We can easily extend this definition to endomorphisms over paths (or essential paths) as follows:
\begin{equation}
    (\eta \otimes \rho)\star(\eta' \otimes \rho')=(\eta \star \eta' ) \otimes (\rho \star \rho'). 
\end{equation}

We will also need to define an involution, called ``the star''. Let $\eta = (v_{0},...,v_{n})$, then $\eta^{\star}=(v_{n},...., v_{0}).$ Notice that $\eta ^\star$ is the path obtained from $\eta$ by ``time inversion''.

\vskip 0.3cm
Extending this definition we can see what is the star of a graded endomorphism $(\eta \otimes \rho $): 
\begin{equation}
    (\eta \otimes \rho)^{\star}=\eta^{\star}\otimes \rho^{\star}.
    \label{def star}
\end{equation}

While we have only used elementary paths for our definitions, these can be extended to any path. Just bear in mind that the concatenation product is linear and the star satisfies these two properties:
\begin{equation}
    (\eta+\rho)^{\star}=\eta^{\star}+\rho^{\star} \; \forall \rho,\eta \in \mathcal{P},
\end{equation}
\begin{equation}
    (\alpha \eta)^{\star}=\overline{\alpha}\eta^{\star} \; \forall \eta \in \mathcal{P} \; \text{and}\; \forall \alpha \in \mathbb{C}.
\end{equation}

\subsection{The Projector}

For the construction of our weak $\star$ bialgebra we need an associative product and a coassociative coproduct which are compatible. One way to achieve these construction is through the computation of the Ocneanu cells, as mentioned in the Introduction (\ref{sec:level1}), nevertheless we choose to follow the ideas presented by Trinchero \cite{Trinchero:2010yr} because they are more likely generalizable to higher levels, as we will see in section \ref{grafos A1}. We start them by defining a certain projector map $P:End^{gr}(\mathcal{P})\rightarrow \mathcal{A}$ . However, to define $P$, we should first look at how to decompose any $\mathcal{P}_{n}$ into a direct sum of subspaces which can be described completely in terms of the essential paths and the creation operators acting on them. For this case ($A_1$), it is easy to identify that:
\begin{gather}
    \mathcal{P}_{0}=\mathcal{E}_{0},\qquad
    \mathcal{P}_{1}=\mathcal{E}_{1},\\
    \mathcal{P}_{2}=T^{\dagger}_{0}(\mathcal{E}_{1})\bigoplus F^{\dagger}_{0}(\mathcal{E}_{0})\bigoplus B^{\dagger}_{0}(\mathcal{E}_{0}).
    \label{descomposicion p2}
\end{gather}
The first two equations are evident from the definition of essential paths. The third equation comes from the fact that the $2$-step paths are either $2$-step loops or $2$ edges of a triangle. Now, notice that the form of the decomposition of $\mathcal{P}_{2}$ (\ref{descomposicion p2}) suggests us the following ansatz:
\begin{equation}
    \mathcal{P}_{n}=T_{0}^{\dagger}(\mathcal{P}_{n-1})\bigoplus F_{0}^{\dagger}(\mathcal{P}_{n-2})\bigoplus B_{0}^{\dagger}(\mathcal{P}_{n-2}).
    \label{decomposition}
\end{equation}

To prove that this decomposition is always possible, we will use induction. So, for now, let us assume that for any $k<n$ the decomposition given by equation (\ref{decomposition}) holds. It is left for us to prove that the proposed decomposition of $\mathcal{P}_n$ has the adequate dimension and that the subspaces involved are orthogonal to each other. By noticing that:
\begin{equation}
    C_{i,j}C_{k,j}^{\dagger}=\beta \delta_{ik},
\end{equation}
we can conclude that $T_{0}^{\dagger}(\mathcal{P}_{n-1})$, $F_{0}^{\dagger}(\mathcal{P}_{n-2})$, and $B_{0}^{\dagger}(\mathcal{P}_{n-2})$ are indeed mutually orthogonal subspaces. On the other hand, the dimension of $\mathcal{P}_{k}$ is always $3\cdot 2^{k}$, which is the number of ways in which we can choose an elementary path with $k$ steps. There are $3$ choices for the first vertex, two choices for the next vertex, and so on. Suppose that we choose $2$ paths $\eta$ and $\rho$ from an orthogonal basis of $\mathcal{P}_{n-1}$. Then, we can see that
\begin{equation}(T^{\dagger}_{0}\eta, T^{\dagger}_{0} \rho)=(T_{0}T^{\dagger}_{0}\eta,\rho)=\beta(\eta,\rho)=0.
\end{equation}

Similarly, we can choose any two orthogonal paths $\lambda$ and $\xi$ $\in$ $\mathcal{P}_{n-2}$ to see that
\begin{gather}
    (F^{\dagger}_{0}\lambda, F^{\dagger}_{0} \xi)=0,\\
    (B^{\dagger}_{0}\lambda, B^{\dagger}_{0} \xi)=0.
\end{gather}
Therefore, if $\{\eta_{i}\}$ is an orthogonal basis of $\mathcal{P}_{n-1}$, then the list $\{T^{\dagger}_{0}(\eta_{i})\}$ is a list of mutually orthogonal paths, which also spans the vector space $T^{\dagger}_0(\mathcal{P}_{n-1})$. We can see this by using the linearity of $T^{\dagger}_{0}$:
\begin{equation}
    T^{\dagger}_{0}\left(\sum_{i}\alpha_{i}\eta_{i}\right)=\sum_{i}\alpha_{i}T^{\dagger}(\eta_{i}).
\end{equation}

As a result, $\{T^{\dagger}_{0}(\eta_{i})\}$ is an orthogonal basis for $T^{\dagger}_{0}(\mathcal{P}_{n-1})$. Likewise, orthogonal bases for $F_0^\dagger(\mathcal{P}_{n-2})$ and $B_0^\dagger(\mathcal{P}_{n-2})$ can be found by simply applying the operators $F^{\dagger}_{0}$ and $B^{\dagger}_{0}$ over the elements of an orthogonal basis for $\mathcal{P}_{n-2}$ respectively. As a consequence, $\mathcal{P}_{n-1}$ and $T^{\dagger}_{0}(\mathcal{P}_{n-1})$ have the same dimension and this is also true for $F^{\dagger}_{0}(\mathcal{P}_{n-2})$ and $B^{\dagger}_{0}(\mathcal{P}_{n-2})$. The dimension of $T^{\dagger}_{0}(\mathcal{P}_{n-1})\bigoplus F^{\dagger}(\mathcal{P}_{n-2}) \bigoplus B^{\dagger}(\mathcal{P}_{n-2})$ is then 
\begin{equation}
3\cdot 2^{n-1}+3\cdot 2^{n-2}+3\cdot 2^{n-2}=3\cdot 2^{n},
\end{equation}
which is actually the dimension of $\mathcal{P}_{n}$. As this decomposition has the right dimension and all the subspaces are orthogonal to each other, we conclude that (\ref{decomposition}) holds. $\blacksquare$

One useful result we get from this decomposition is 
\begin{equation}
\begin{split}
    (C_{i_{n},0}^{\dagger}...C_{i_{1},0}^{\dagger}\xi,C_{j_{m},0}^{\dagger}...C_{j_{1},0}^{\dagger}\omega) &= \beta^{n}\delta_{nm}\delta_{i_{n}j_{m}}....\delta_{i_{1}j_{1}}\delta_{\xi,\omega} \\ &= C(j_1^0, ..., j_m^0; i_n^0, ..., i_1^0).
    \end{split}
\end{equation}
So, now that we have the decomposition, we are ready to define the projector $P:End^{gr}(\mathcal{P})\rightarrow \mathcal{A}$ as:

\begin{equation}
\begin{split}
    P(C_{i_{n},0}^{\dagger}...C_{i_{1},0}^{\dagger}\xi & \otimes C_{j_{m},0}^{\dagger}...C_{j_{1},0}^{\dagger}\omega)\\ & =\sum_{\rho\in\mathcal{E}}(C_{j_{1},0}...C_{j_{m},0}C_{i_{n},0}^{\dagger}...C_{i_{1},0}^{\dagger}\xi,\rho)\rho \otimes \omega\\
    &=\beta^{n}\delta_{nm}\delta_{i_{n}j_{m}}....\delta_{i_{1}j_{1}}\xi \otimes \omega. 
    \end{split}
\end{equation}
The sum above requires that $\rho$ has the same length as $\omega$ and $\xi$. Notice that, any path is a linear combination of terms of the form $C_{i_{n},0}^{\dagger}...C_{i_{1},0}^{\dagger}\xi$, which are orthogonal. As all our operators are linear, we only need to make the proofs in the rest of the work for those terms. 

\subsection{The Product}

Using the projector of the last section, we can define a bilinear form $(\cdot): \mathcal{A} \times \mathcal{A}\rightarrow \mathcal{A}$, which will be called ``the product'' of $\mathcal{A}$
\begin{equation}
(\eta \otimes \eta')\cdot (\rho \otimes \rho')=P(\eta\star\rho \otimes \eta'\star\rho') \quad \forall \eta, \, \eta' \, \rho \; \mbox{and} \; \rho' \, \in \mathcal{E}
\label{definition of the product}
\end{equation}

The identity element associated with this product is simply: 

\begin{equation}
    \mathbb{1}= \sum_{v,u \in \mathcal{E}_0} v \otimes u
    \label{identity of the product}
\end{equation}

To write the corresponding multiplication table we should do the following. We label the vertices $(1)$, $(3)$, and $(\overline{3})$ as $z_0$, $z_1$, and $z_2$ respectively. Now, we label the elementary paths of length $(1,0)$, which are $(13)$, $(3\overline{3})$, and $(\overline{3}1)$ as $x_0$, $x_1$, and $x_2$ respectively. Finally, the elementary paths of length $(0,1)$ which are $(31)$, $(\overline{3}3)$ and $(1\overline{3})$ are labeled as $y_0$, $y_1$, and $y_2$ respectively. Notice that these paths constitute an orthogonal basis of $\mathcal{E}$ and therefore, we can write any graded endomorphism in $\mathcal{A}$ in terms of these paths.

\begin{table*}
\begin{ruledtabular}
  \caption{  
    \label{tab:my_label} Multiplication table for the elements of the algebra. For simplicity, it is assumed that before evaluating these expressions, one must take the $\pmod{3}$ of each index.}
    \begin{tabular}{cccc}
  
        $\cdot$ & $z_i \otimes z_j$ & $x_i \otimes x_j$ & $y_i \otimes y_j$  \\ \hline
       $z_l \otimes z_m$  & $\delta_{li} \delta_{mj} (z_l \otimes z_m)$ & $\delta_{li} \delta_{mj}(x_i \otimes x_j)$ & $\delta_{(l+2)i} \delta_{(m+2)j}(y_i \otimes y_j)$\\
        
       $x_l \otimes x_m$ & $\delta_{l(i+2)} \delta_{m(j+2)}(x_l \otimes x_m)$ & $\delta_{l(i+2)} \delta_{m(j+2)}(y_{i+1}\otimes y_{j+1})$ & $\delta_{li} \delta_{mj} (z_i \otimes z_j)$\\ 
     
       $y_l \otimes y_m$ & $\delta_{li} \delta_{mj} (y_l \otimes y_m)$ & $\delta_{li} \delta_{mj} (z_{i+1} \otimes z_{j+1})$ & $\delta_{(l+2)i}\delta_{(m+2)j}(x_{l+1} \otimes x_{m+1})$ \\ 
  
    \end{tabular}
    \end{ruledtabular}
\end{table*}

Notice that, in Hammaoui's work \cite{HammahuiA2SU32008} it is defined a very similar product for this case, using the Ocneanu cells method. However, according to the multiplication table presented in that work, 
\begin{equation}
(x_{i}\otimes x_{j}) \cdot (y_{k}\otimes y_{l})=(y_{i}\otimes y_{j}) \cdot (x_{k}\otimes x_{l})=0.
\label{wrong}
\end{equation}
But, in fact, this leads to a \textbf{non-associative product}, because then we would have
\begin{equation}
    [(x_{i}\otimes x_{j})\cdot (y_{i} \otimes y_{j})] \cdot (y_{i+2}\otimes y_{j+2})=0,\end{equation}
but
\begin{equation}
    (x_{i}\otimes x_{j})\cdot [(y_{i} \otimes y_{j}) \cdot (y_{i+2}\otimes y_{j+2})]=y_{i+2}\otimes y_{j+2}.
\end{equation}
 Instead, the definition of the product given in Table \ref{tab:my_label} is compatible with the $\star$ and associative. This means that, for $a=\xi_{1}\otimes \xi_{1}',$ $b=\xi_{2}\otimes \xi_{2}',$ and $c=\xi_{3}\otimes \xi_{3}',$ then 

\begin{gather}
      (a \cdot b)^\star = (b)^\star \cdot (a)^\star, 
      \label{antihomomorphism of the star}\\
      \begin{split}
       (a \cdot b)  \cdot c= a\cdot (b \cdot c). 
       \label{associativity of the product}
       \end{split}
\end{gather}

In order to prove the anti-homomorphism property of the involution with the product (\ref{antihomomorphism of the star}) an intermediate result is needed:
\begin{equation}
P((\eta \otimes \eta')^\star)= (P(\eta \otimes \eta'))^\star \qquad \forall \; \; \eta , \eta ' \; \in \mathcal{P}_{n}.
\label{la estrella de la proyeccion}
\end{equation}
It is easy to see that (\ref{la estrella de la proyeccion}) is trivially satisfied if $P(\eta \otimes \eta')=0$. In case it does not vanish take $\eta=C^{\dagger}_{i_{n},0}...C^{\dagger}_{i_{1},0}\xi_{a}$ and $\eta'=C^{\dagger}_{i_{n},0}...C^{\dagger}_{i_{1},0}\xi_{a}$, and remembering that $(\eta^{\star},\eta'^{\star})=\overline{\eta, \eta'}$ you can immediately notice that
\begin{equation}
    \begin{split}
    P(\eta \otimes \eta') &= \sum_{\xi_{c}\in \mathcal{E}}(\eta,C^{\dagger}_{i_{n},0}...C^{\dagger}_{i_{1},0}\xi_{c}) \xi_{c} \otimes \xi_{b}\\
                                  &= \sum_{\xi_{c}\in \mathcal{E}}\overline{(\eta^{\star},(C^{\dagger}_{i_{n},0}...C^{\dagger}_{i_{1},0}\xi_{c})^{\star})} \xi_{c} \otimes \xi_{b}.
\end{split}
\end{equation}
Now, we can use that in this sum the only term which survives is the one for which $\xi_{c}=\xi_{b}$, if any. Thus
\begin{equation}
    P(\eta \otimes \eta') = \sum_{\xi_{c}\in \mathcal{E}}\overline{(\eta^{\star },\eta'^{\star })} \xi_{c} \otimes \xi_{b}.
\end{equation}
If we take the star for the last equation, equation (\ref{la estrella de la proyeccion}) follows $\blacksquare$. Proving equation (\ref{antihomomorphism of the star}) is done by making straightforward calculations and using the property (\ref{la estrella de la proyeccion}). 




Now, the associativity is a bit harder to prove. To do so, we must first prove the following result: $ \forall \, \xi, \xi' \in \mathcal{E} \; \mbox{and} \; \forall \,\eta, \eta' \in \mathcal{P}$ then
{\small
\begin{gather}
    P((\xi \otimes \xi') \star P(\eta \otimes \eta'))= P((\xi \otimes \xi') \star (\eta \otimes \eta')),
    \label{proposition 29 (a)}\\
    P(P(\eta \otimes \eta') \star (\xi \otimes \xi')) = P((\eta \otimes \eta') \star (\xi \otimes \xi')),
    \label{proposition 29 (b)}
\end{gather}}
First, we can notice that both of these equations are trivially satisfied if $P(\eta \otimes \eta')=0$. 

 Supposing that this does not happen,  we can start by looking at the left-hand side of equation (\ref{proposition 29 (a)}) and considering paths $\eta=C^{\dagger}_{i_{n}}...C^{\dagger}_{i_{1}}\xi_{a}$ and $\eta'=C^{\dagger}_{i_{n}}...C^{\dagger}_{i_{1}}\xi_{b}$. In that case, $$P(\eta \otimes \eta')=\beta^{n}\xi_{a} \otimes \xi_{b},$$ and the lhs of equation (\ref{proposition 29 (a)}) is $$P((\xi \otimes \xi') \star P(\eta \otimes \eta'))= \beta^n P((\xi \otimes \xi')\star (\xi_a \otimes \xi_b)).$$

On the other hand, the rhs of equation (\ref{proposition 29 (a)}) is
{\small
\begin{equation}
    \begin{split}
    P((\xi\otimes \xi') \star(\eta \otimes \eta'))&= P(\xi \star C^\dagger_{i_n,0}... C^\dagger_{i_1,0} \xi_a \otimes \xi' \star C^\dagger_{i_n,0}... C^\dagger_{i_1,0} \xi_b)\\ &= P(C^\dagger_{i_n,l}...C^\dagger_{i_1,l}(\xi \star \xi_a) \otimes C^\dagger_{i_n,l}...C^\dagger_{i_1,l} (\xi' \star \xi_b))\\ &= \beta^{n}P((\xi \star \xi_a) \otimes (\xi' \star \xi_b)),
    \label{rhs proposition 29}
    \end{split}
\end{equation}}
which comes from three facts. First, that $$\rho \star C^\dagger_{i_n,0}... C^\dagger_{i_1,0} \rho' = C^\dagger_{i_n,l}...C^\dagger_{i_1,l}(\rho \star \rho'),$$ 
for $\rho$ and $\rho'\in \mathcal{P}$ and $l$ being the number of steps of $\rho$. The second and third facts are the definition of the projector and the inner product of the space of paths. Thus, both sides of equation (\ref{proposition 29 (a)}) are equal, in general $\blacksquare$. The proof for the equation (\ref{proposition 29 (b)}) is quite similar. As a direct consequence of relations (\ref{proposition 29 (a)}) and (\ref{proposition 29 (b)}), the associativity follows: let $a=\xi_{1}\otimes \xi_{1}',$ $b=\xi_{2}\otimes \xi_{2}',$ and $c=\xi_{3}\otimes \xi_{3}',$ then 

\begin{equation}
    \begin{split}
         (a \cdot b) \cdot c &=P(P(a \star b) \star c)\\
         &=P(a\star b \star c)\\
         &= P(a \star P(b \star c))\\
         &= a\cdot (b \cdot c) \quad \blacksquare. 
    \end{split}
\end{equation}

\subsection{The Co-product}

The coproduct $\Delta$ is a linear map $\Delta: \mathcal{A} \rightarrow \mathcal{A} \times \mathcal{A} $, which we will define as: 

\begin{equation}
    \Delta(\xi \otimes \xi')= \sum_{\substack{\xi_a \in \mathcal{E} \\ \#\xi_a=\# \xi}} \xi \otimes \xi_a \boxtimes \xi_a \otimes \xi', 
    \label{definition coproduct}
\end{equation}

where the sum runs over a complete orthonormal basis of $\mathcal{E}$ and $\# \xi$ denotes the length of $\xi.$ Next, by straightforward calculations, it can be proved that this coproduct is compatible with the star 
\begin{equation}
    \Delta(a^\star)= \Delta(a)^\star.
    \label{coproduct 2}
\end{equation}

In addition, the coproduct is co-associative:
 
 \begin{equation}
     \begin{split}
         (\Delta \otimes Id) \Delta(\xi \otimes \omega) &= (\Delta \otimes Id)\sum_{\rho \in \mathcal{E}} \xi \otimes \rho \boxtimes \rho \otimes \omega \\ 
         &= \sum_{\rho, \rho' \in \mathcal{E}} \xi \otimes \rho' \boxtimes \rho' \otimes \rho \boxtimes \rho \otimes \omega \\ 
         &= (Id \otimes \Delta) \left(\sum_{\rho' \in \mathcal{E}} \xi \otimes \rho' \boxtimes \rho' \otimes \omega\right) \\
         &= (Id \otimes \Delta) \Delta(\xi\otimes \omega)\quad \blacksquare. 
     \end{split}
 \end{equation}

    

Another interesting property is that
\begin{equation}
    \Delta P= P^{\otimes 2} \Delta_{\mathcal{P}},
    \label{proposition coproduct 1}
\end{equation}

where $P^{\otimes 2}$ stands for two copies of the projector, and

\begin{equation}
    \Delta_{\mathcal{P}} (\chi \otimes \chi')= \sum_{\eta \in \mathcal{P}}\chi \otimes \eta \boxtimes \eta \otimes \chi'.
\end{equation}

Here the sum runs over a complete orthonormal basis of $\mathcal{P}.$ To prove this, notice that equation (\ref{proposition coproduct 1}) is trivially satisfied if $P(\eta \otimes \eta')=0$. So, let's suppose that this does not happen and have a look at the lhs of equation (\ref{proposition coproduct 1}). If $\eta =C^\dagger_{i_n,0} ... C^\dagger_{i_1,0} \xi_a$ and $\eta'= C^\dagger_{i_n,0} ... C^\dagger_{i_1,0} \xi_b$, then $P(\eta \otimes \eta')=\beta^{n} \xi_{a}\otimes \xi_{b}$. As a consequence it follows that

\begin{equation}
    \begin{split}
        \Delta P(\eta \otimes \eta')&= \Delta[ \beta^{n} \xi_a \otimes \xi_b]= \beta^{n} \Delta(\xi_a \otimes \xi_b)\\
        &= \beta^{n} \sum_{\substack{\xi_c \in \mathcal{E}\\ \# \xi
        _c = \# \xi_a}} \xi_a \otimes \xi_c \boxtimes \xi_c \otimes 
        \xi_b.
    \end{split}
\end{equation}

In the same way, the rhs of equation (\ref{proposition coproduct 1}) is: 

\begin{equation}
    \begin{split}
        P^{\otimes 2} \Delta_\mathcal{P}(\eta \otimes \eta') &= P^{\otimes 2} \sum_{\rho \in \mathcal{P}} \eta\otimes \rho \boxtimes \rho \otimes \eta' \\&= \sum_{\rho \in \mathcal{P}}P(\eta \otimes \rho) \boxtimes P(\rho \otimes \eta')\\
        &= \beta^{n}\sum_{\substack{\xi_c \in \mathcal{E} \\ \#\xi_a = \#\xi_c}}  \xi_a \otimes \xi_c \boxtimes \xi_c \otimes \xi_b,
    \end{split}
\end{equation}
where we must notice that the only $\rho$'s surviving the sum are of the form $\rho=\beta^{-n/2} C^\dagger_{i_n,0} ... C^\dagger_{i_1,0} \xi_{c}$ with $\xi_{c}\in \mathcal{E}$. Therefore, equation (\ref{proposition coproduct 1}) holds, in general $\blacksquare$.

Next, we can also prove the following property:

\begin{equation}
\begin{split}
    P^{\otimes2} (\Delta_{\mathcal{P}}(\xi_a \otimes \xi_b)& \star \Delta_{\mathcal{P}}(\xi_c \otimes \xi_d))= \\ &P^{\otimes 2}[P^{\otimes 2}\Delta_{\mathcal{P}}(\xi_a\otimes \xi_b) \star P^{\otimes 2}\Delta_{\mathcal{P}}(\xi_c \otimes \xi_d)].
    \label{proposition coproduct 2}
    \end{split}
\end{equation}

To see that this is actually true, we first realize that, for any $(\rho \otimes \eta)$ $\in$ $\mathcal{A}$:
\begin{equation}
    \Delta P (\rho \otimes \eta)=P^{\otimes 2}\Delta_{\mathcal{P}}(\rho \otimes \eta)=P^{\otimes 2}\Delta (\rho \otimes \eta),
\end{equation}
where we have used equation (\ref{proposition coproduct 1}). Then, equation (\ref{proposition coproduct 2}) holds trivially $\blacksquare$ .

Using equations (\ref{proposition coproduct 1}) and (\ref{proposition coproduct 2}), the compatibility with the product follows 
{\small
\begin{equation}
     \begin{split}
         \Delta((\xi \otimes \xi') \cdot (\rho \otimes \rho'))&= \Delta[P((\xi \otimes \xi') \star (\rho \otimes \rho'))]\\
         &= P^{\otimes 2} \Delta_{\mathcal{P}} ((\xi \otimes \xi') \star (\rho \otimes \rho'))\\
         &= P^{\otimes 2}[\Delta_{\mathcal{P}} (\xi \otimes \xi') \star \Delta_{\mathcal{P}}(\rho \otimes \rho')]\\ 
         &= P^{\otimes 2}[P^{\otimes 2} \Delta_{\mathcal{P}} (\xi \otimes \xi') \star P^{\otimes 2} \Delta_{\mathcal{P}} (\rho \otimes \rho')]\\
         &= P^{\otimes 2}[\Delta(P(\xi \otimes \xi')) \star \Delta (P(\rho\otimes \rho'))]\\ 
         &= \Delta(\xi \otimes \xi') \cdot \Delta(\rho \otimes \rho')\quad\blacksquare.
     \end{split}
 \end{equation}}
\subsection{The unit and the co-unit}
The unit is a map $\sigma: \mathbb{C} \rightarrow \mathcal{A}$  defined as: 
\begin{equation}
    \sigma(c) = c \mathbb{1} \quad \forall c \in \mathbb{C}.
\end{equation}
This unit map is trivially compatible with the product previously defined. On the other hand, the counit $\epsilon$ is a map $\epsilon: \mathcal{A}  \rightarrow \mathbb{C} $ and, in this case, is defined as:
\begin{equation}
    \epsilon(\xi \otimes \xi')= (\xi, \xi').
    \label{definition of the co-unit}
\end{equation}
Given the definition of the inner product, it follows that for every $a$ $\in$ $\mathcal{A}$ \begin{equation}
\epsilon(aa^{\star })\geq 0.
\end{equation}
 It is also straightforward that 
\begin{equation}
    (\epsilon \otimes Id) \Delta = Id = (Id \otimes \epsilon) \Delta. \label{coassociativity of the co-unit}
\end{equation}
Another interesting property is that, for every $\eta, \eta' \; \in \; \mathcal{P}$: 
\begin{equation}
    \epsilon (P(\eta \otimes \eta'))=\epsilon(\eta \otimes \eta').
\end{equation}
This equation is trivially satisfied when $P(\eta \otimes \eta')=0$, because both sides vanish. When $P(\eta \otimes \eta')$ does not vanish, we have
\begin{equation}
    \epsilon (P(\eta \otimes \eta'))=\beta^{n}(\xi_{a}, \xi_{b})=\epsilon(\eta \otimes \eta'),
\end{equation}
where $\eta=C^{\dagger}_{i_{n},0}...C^{\dagger}_{i_{1},0}\xi_{a}$ and $\eta'=C^{\dagger}_{i_{n},0}...C^{\dagger}_{i_{1},0}\xi_{b}$ $\blacksquare$.
\vskip 0.2cm
Finally, the co-unit is also compatible with the product previously defined: 
\begin{equation}
       \epsilon(ab)= \epsilon(a \mathbb{1}_1)\epsilon(\mathbb{1}_2 b) \quad \forall a,b \in \mathcal{A}, \label{compatibility of the co-unit with the product}
\end{equation}
where the Sweedler convention is used. So, using the definition of the identity element of the product (\ref{identity of the product}), then 
\begin{equation}
    \begin{split}
        \Delta(\mathbb{1}) &= \sum_{v,u,\rho \in \epsilon_0} u\otimes \rho \boxtimes \rho  \otimes v\\
        &=\sum_{\rho \in \epsilon_0} \left( \sum_{u \in \epsilon_0} u\otimes \rho \boxtimes \sum_{v \in \epsilon_0} \rho \otimes v \right)\\ 
        &= \mathbb{1}_1 \boxtimes \mathbb{1}_2.
    \end{split}
\end{equation}
To show that equation (\ref{compatibility of the co-unit with the product}) is true we start by computing the lhs 
\begin{equation}
    \begin{split}
        \epsilon(\xi \otimes \omega \cdot \xi' \otimes \omega')&= \epsilon(P( \xi \star \xi' \otimes \omega \star \omega'))\\ &= (\xi \star \xi', \omega \star \omega'),
    \end{split}
\end{equation}
while the rhs is 
\begin{equation}
    \begin{split}
         \epsilon(\xi& \otimes \omega \cdot \mathbb{1}_1) \epsilon( \mathbb{1}_2  \cdot \xi' \otimes \omega')\\&= \sum_{u,v, \rho \in \mathcal{E}_0} \epsilon(\xi \otimes \omega \cdot v\otimes u) \epsilon( u\otimes \rho \cdot \xi' \otimes \omega')\\ 
         &= \sum_{u,v, \rho \in \mathcal{E}_0} \delta_{r(\xi)v} \delta_{r(\omega)u} (\xi, \omega) \delta_{l(\xi')u} \delta_{l(\omega')\rho}(\xi', \omega')\\ 
         &= \delta_{r(\xi)l(\xi')}\delta_{r(\omega)l(\omega')} (\xi, \omega)(\xi', \omega')\\
         &= (\xi \star \xi', \omega \star \omega').
    \end{split}
\end{equation}
where $r(\eta)$ and $l(\eta)$ denote the last and the first vertices of $\eta$, respectively. Therefore, equation (\ref{compatibility of the co-unit with the product}) holds $\blacksquare$. 
\subsection{The Antipode} 
 The antipode is a map $S:\mathcal{A} \rightarrow \mathcal{A}$ defined by: 
 \begin{equation}
      S(\xi,\omega)=\omega^{\star}\otimes \xi^{\star}.
      \label{def antipode}
\end{equation}
This definition is motivated by the ansatz of Trinchero \cite{Trinchero:2010yr}. Using equations (\ref{def antipode}) and (\ref{def star}) it is fairly easy to prove that the antipode is compatible with the star:
\begin{equation}
    S[(S(a^\star))^\star]= a. \label{compatibility of the antipode with the star}
\end{equation}
It is also straightforward to prove that the antipode is compatible with the coproduct 
\begin{equation}
    \Delta(S(a))=S \otimes S (\Delta^{op}(a)), \label{compatibility of the antipode with the coproduct}
\end{equation}
where 
\begin{equation}
       \Delta ^{op}(\xi\otimes \omega)= \sum_{\rho \in \epsilon} \rho \otimes \omega \boxtimes \xi \otimes \rho. 
\end{equation}
If the projector does not vanish consider $\eta,\;\eta',\;\rho, \; \rho' \in \mathcal{E}$ such that $\eta \star \rho=\beta^{-n/2}C^\dagger_{i_n,0}... C^\dagger_{i_1,0}\xi_a$ and $\eta' \star \rho' = \beta^{-n/2}C^\dagger_{i_n,0} ... C^\dagger_{i_1,0}\xi_b$, so that $P(\eta \star \rho \otimes \eta' \star \rho')=\xi_{a}\otimes \xi_{b}$. 
\vskip 0.2cm
Then, by straightforward calculations
\begin{equation}
\begin{split}
    S(\eta \otimes \eta'\cdot \rho \otimes \rho') &= S(P(\eta \star \rho \otimes \eta' \star \rho'))\\&=S(\xi_a \otimes \xi_b)\\&=\xi_b^\star \otimes \xi_a^\star.
\end{split}    
\end{equation}
On the other hand, by equation (\ref{antihomomorphism of the star}) and the definition of the antipode
\begin{equation}
    \begin{split}
        S(\rho \otimes \rho') \cdot S(\eta \otimes \eta') &= (\rho'^\star \otimes \rho^\star) \cdot (\eta'^\star \otimes \eta^\star)\\ &= \left[ (\eta' \otimes \eta)\cdot (\rho' \otimes \rho)\right]^{\star}\\ &= \xi_{b}^{\star}\otimes \xi_{a}^{\star}.
    \end{split}
\end{equation}
Therefore, it follows that the antipode is also compatible with the product 
\begin{equation}
    S(ab) =S(b) S(a). \label{compatibility of the antipode with the product}
\end{equation}
Another property for the antipode is 
\begin{equation}
\begin{split}
    \sum_{\xi',\omega' \in \mathcal{E}}S(\xi\otimes& \xi') \cdot (\xi' \otimes \omega') \boxtimes \omega' \otimes \omega\\&= \sum_{u,v,v'\in \mathcal{E}_{0}}u\otimes v \boxtimes (\xi \otimes \omega)\cdot (v\otimes v').
    \label{weird property antipode}
    \end{split}
\end{equation}
In order to prove this property, we start by computing the rhs of equation (\ref{weird property antipode}) 
\begin{equation}
\begin{split}
    \sum_{u,v,v'\in \mathcal{E}_{0}}u\otimes v &\boxtimes (\xi \otimes \omega)\cdot (v\otimes v')\\&=\sum_{u,v,v'\in \mathcal{E}_{0}}u\otimes v \boxtimes (\xi \otimes \omega)\delta_{r(\xi),v}\delta_{r(\omega),v'}\\
&=\sum_{u\in \mathcal{E}_{0}}u\otimes r(\xi) \boxtimes (\xi \otimes \omega).
\end{split}
\end{equation}
In the same way, the lhs of eqaution (\ref{weird property antipode}) is
\begin{equation}
\begin{split}    \sum_{\xi',\omega'\in \mathcal{E}}S(\xi\otimes \xi') &\cdot (\xi' \otimes \omega')  \boxtimes \omega' \otimes \omega\\ &=\sum_{\xi',\omega'\in \mathcal{E}}(\xi'^{\star }\otimes \xi^{\star }) \cdot (\xi' \otimes \omega') \boxtimes \omega' \otimes \omega\\
    &=\sum_{\xi',\omega'\in \mathcal{E}}P(\xi'^{\star }\star \xi'\otimes \xi^{\star }\star \omega')\boxtimes \omega' \otimes \omega.
\end{split}
\end{equation}
At this point, notice that if $\omega' \neq \xi  $ then $P(\xi'^{\star }\star \xi'\otimes \xi^{\star }\star \omega')$ vanishes. The path $\xi$ has $1$ step at most, so $(\xi'^{\star}\star \xi')$ is a $2$-step loop or a vertex. Therefore, $\xi^{\star}\star \omega'$ should also be a $2$-step loop with the same length or a vertex, otherwise, the projector would vanish. In any of these cases, it follows that
\begin{equation}
\begin{split}
    \sum_{\xi',\omega'\in \mathcal{E}}S(\xi\otimes \xi') \cdot (&\xi' \otimes \omega') \boxtimes \omega' \otimes \omega\\
    &=\sum_{\xi'\in \mathcal{E}}P(\xi'^{\star }\star \xi'\otimes \xi^{\star }\star \xi)\boxtimes \xi \otimes \omega\\
    &=\sum_{u\in \mathcal{E}}u\otimes r(\xi)\boxtimes \xi \otimes \omega,
    \end{split}
\end{equation}
where it is necessary to notice that there are exactly three possible choices for $\xi'$ which have the same length of $\xi$ (and therefore, which contribute to the sum), each of which have a different starting vertex. That is why we have changed the sum over $\xi'$ by a sum over the starting vertex $u$ of $\xi'$. As a consequence, equation (\ref{weird property antipode}) holds $\blacksquare$.

\section{About the $SU(N)$ $A_1$ graphs}
\label{grafos A1}

So far, in all the proofs provided in this work, for the construction of the quantum groupoid associated to $A_1$ of $SU(3)$, the only assumptions we have made are that essential paths are paths with $1$ or $0$ steps and that the length is an ordered pair indicating the number of times that each generator is applied to obtain the corresponding path. Moreover, the key ingredient in our construction is that we can actually achieve the decomposition of any $\mathcal{P}_{n}$ in terms of essential paths and creation operators. As a consequence, our previous results hold if we manage to find the decomposition and if we properly define the length of a path (in terms of the number of generators) and the associated essential paths for the graph under study, regardless of its topology. 

To show how to obtain the decomposition of any $A_{1}$ $SU(N)$ graph, we are going to explicitly show one more complex example, which is, the decomposition for $\mathcal{P}_{n}$ of the $A_{1}$ $SU(4)$ graph. The three associated graphs (one for each generator of $SU(4)$) are shown below.

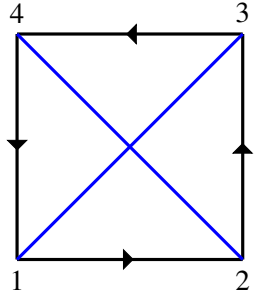
\begin{figure}[H]
\centering
\begin{tikzpicture}
\begin{scope}[very thick, every node/.style={sloped,allow upside down}]
  \draw (0,0) node[below] {1}-- node {\midarrow} (3,0);
  \draw (3,0) node[below] {2}-- node {\midarrow} (3,3) ;
  \draw (3,3) node[above] {3} -- node {\midarrow} (0,3);
  \draw (0,3) node[above] {4} -- node {\midarrow} (0,0);
  \draw[blue] (0,0) --  (3,3); 
  \draw[blue] (0,3) --  (3,0);  
\end{scope}
\end{tikzpicture}
\caption{\label{a1 su4}The $A_1$ $SU(4)$ graph. $SU(4)$ has $3$ fundamental generators which will be called $\sigma_1$, its conjugate $\overline{\sigma}_1$ and $\sigma_2$ ($\overline{\sigma}_2= \sigma_2$). In this diagram $\sigma_1$ is represented with black oriented segments and $\sigma_2$ with blue segments \cite{Robertpage}. }
\end{figure}

This is a complete graph, which means that every pair of vertices is connected by an edge. In addition, note that each edge belongs to two different triangle-like paths which connects the same two vertices in the same order. Also, from each vertex we can find $3$ different $2$-step loops which start at that vertex. Each one of these paths correspond to a way in which two of the generators of the irreps of $SU(4)$ are applied. In fact, there are $9$ ways to choose the pair of generators. Three of them correspond to the $2$-step loops, which are $\sigma_{1}\overline{\sigma}_{1},$ $\overline{\sigma}_{1}\sigma_{1},$ and $\sigma_2 \sigma_2.$ Notice that the rightmost generator is applied first and then we apply the other generator. We can define one operator for each of these back-tracks, which are clearly different given that they are associated to a different combination of the generators. So, let's call them $L^\dagger_{0,i},$ $L^\dagger_{1,i}$ and $L^\dagger_{2,i}.$ 

With the triangle-like paths, we have $6$ associated combinations. However, as there are only two associated triangles for each oriented edge, we can summarize these combinations into $2$ operators as follows. 

\begin{equation}
    T^\dagger_{0,i} \rightarrow 
    \begin{cases}
    \sigma_1 \sigma_1 \quad \text{when applied over an edge in $\sigma_2$} \\
    \sigma_2 \overline{\sigma}_1 \quad \text{when applied over an edge in $\sigma_1$}\\
    \sigma_2 \sigma_1 \quad \text{when applied over an edge in $\overline{\sigma}_1$}
    \end{cases}
\end{equation}
\begin{equation}
T^\dagger_{1,i}\rightarrow
    \begin{cases}
    \overline{\sigma}_1 \overline{\sigma}_1 \quad \text{when applied over an edge in $\sigma_2$}\\
    \overline{\sigma}_1 \sigma_2 \quad \text{when applied over an edge in $\sigma_1$} \\ 
    \sigma_1 \sigma_2 \quad \text{when applied over an edge in $\overline{\sigma}_1$}.
    \end{cases}
\end{equation}

It follows that the decomposition for $\mathcal{P}_2$ of the $A_{1}$ $SU(4)$ graph is then

\begin{equation}
    \mathcal{P}_2= \bigoplus_{i=0}^{2}L^\dagger_{i,0}(\mathcal{E}_0) \bigoplus_{j=0}^{1} T^\dagger_{j,0}(\mathcal{E}_1).
\end{equation}

And recursively, it can be shown that

\begin{equation}
    \mathcal{P}_n= \bigoplus_{i=0}^{2}L^\dagger_{i,0}(\mathcal{P}_{n-2}) \bigoplus_{j=0}^{1} T^\dagger_{j,0}(\mathcal{P}_{n-1}).
\end{equation}

As can be seen, we do not actually need to know how are the generators of the graph to construct the operators for an arbitrary $SU(N)$, given that the $A_{1}$ graphs are always complete. This is because at each vertex we must have $N-1$ different oriented edges going out of that vertex. Also, we must have $N-2$ different triangles associated to each oriented edge. Therefore, we will have $N-1$ loop-like operators and $N-2$ triangle-like operators. It follows that the decomposition for $\mathcal{P}_{2}$ of the $A_{1}$ $SU(N)$ graph  is 

\begin{equation}
    \mathcal{P}_2= \bigoplus_{i=0}^{N-1}L^\dagger_{i,0}(\mathcal{E}_0) \bigoplus_{j=0}^{N-2} T^\dagger_{j,0}(\mathcal{E}_1),
\end{equation}

and recursively, it can be shown that

\begin{equation}
    \mathcal{P}_n= \bigoplus_{i=0}^{N-1}L^\dagger_{i,0}(\mathcal{P}_{n-2}) \bigoplus_{j=0}^{N-2} T^\dagger_{j,0}(\mathcal{P}_{n-1}).
    \label{decomposition general}
\end{equation}

We can show equation (\ref{decomposition general}) in the same way as we did for the $A_{1}$ $SU(3)$ graph using induction. First, we notice that the number of elementary paths with $n$ steps in a complete graph with $N$ vertices is $N(N-1)^{n}.$ Also, by constructing the corresponding annihilation operators it is easy to see that

\begin{equation}
    C_{i,j}C^{\dagger}_{k,j}=\beta \delta_{i,k}
\label{hola}
\end{equation}
where we have used the notation $C_{i,j}$ to denote the corresponding operators. As in $SU(3)$ the $i$ stands for the type of operator and the $j$ stands for the step in which it is applied. In the same way, $\beta$ will be a constant associated to each graph. 

Using equation (\ref{hola}) we can see that the subspaces in the direct sum of equation (\ref{decomposition general}) are indeed orthogonal. In addition, notice that the dimension of $\mathcal{P}_{n-1}$ is $N(N-1)^{n-1}$ and the dimension of $\mathcal{P}_{n-2}$ is $N(N-1)^{n-2}.$ Therefore, the dimension of the decomposition we propose is
\begin{equation}N(N-1)^{n-1}(N-2)+N(N-1)^{n-2}(N-1)=N(N-1)^{n}.\end{equation}

Thus, by induction, the decomposition of $\mathcal{P}_{n}$ holds for arbitrary $n\;\blacksquare.$  

Now, in order for us to define the space of graded endomorphisms (and therefore the whole bialgebra) for any $SU(N)$, the length of an arbitrary path will be given by a $(N-1)$-tuple where each entry corresponds to the number of edges of the path associated to each generator of the graph. 


\begin{figure}[H]
\centering
\begin{tikzpicture}
\begin{scope}[very thick, every node/.style={sloped,allow upside down}]
  \draw (0,1.5) node[above] {5}-- node {\midarrow} (-1.43,0.47);
  \draw (-1.43,0.47) node[left] {1} -- node {\midarrow} (-0.89,-1.22);
  \draw (-0.89,-1.22) node[below] {2} -- node {\midarrow} (0.89,-1.22);
  \draw (0.89,-1.22) node[below] {3} -- node{\midarrow} (1.43,0.46);
  \draw (1.43,0.46) node[right] {4} -- node {\midarrow} (0,1.5);
 \draw[blue] (-1.43,0.46) -- node{\midarrow} (0.89,-1.22);
 \draw[blue] (0.89,-1.22) -- node {\midarrow} (0,1.5);
 \draw[blue] (0,1.5) -- node{\midarrow} (-0.89,-1.22);
 \draw[blue] (-0.89,-1.22) -- node {\midarrow} (1.43,0.46);
 \draw[blue] (1.43,0.46) -- node {\midarrow} (-1.43,0.46);
\end{scope}
\end{tikzpicture}
\caption{\label{pentagono}The $A_1$ $SU(5)$ graph. $SU(5)$ has $4$ fundamental generators, one corresponding to the black oriented segments (and its conjugate), and one corresponding to the blue oriented segments (and its conjugate) \cite{Robertpage}. }
\end{figure}
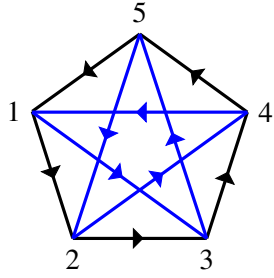
As an example, the table \ref{tab: a1 su4} summarizes the corresponding product associated with the $A_1$ $SU(4)$ graph.

The notation we have used to construct the table is the following. For reference, see figure (\ref{a1 su4}). The vertices have been labeled as $z_0$, $z_1,$ $z_2$ and $z_3$ in counter-clockwise order, starting from $(1).$ Then, the edges associated to $\sigma_1$ have been labeled as $x^{(1)}_i$ in counter clockwise order starting from $(12).$ The edges associated to $\overline{\sigma}_1$ were labeled as $y^{(1)}_i$ and finally, the edges associated with $\sigma_2$ were labeled as $x^{(2)}_i,$ both in counter clockwise order starting from $(21)$ and $(13),$ respectively.

Note that, in general, the product in any $SU(N)$ $A_{1}$ graph can be summarized as follows. Suppose that $(v_{0}v_{1}v_{2})$ and $(v'_{0}v'_{1}v'_{2})$ are two triangle-like paths of the same type, then
\begin{equation}
    (v_{0}v_{1}\otimes v_{0}'v_{1}')\cdot(v_{1}v_{2}\otimes v_{1}'v_{2}')=v_{0}v_{2}\otimes v_{0}'v_{2}'.
\end{equation}
On the other hand, if $v_{0}v_{1}v_{0}$ and $v'_{0}v'_{1}v'_{0}$ are two-step loops of the same type, then
\begin{equation}
    (v_{0}v_{1}\otimes v_{0}'v_{1}')\cdot(v_{1}v_{0}\otimes v_{1}'v_{0}')=v_{0}\otimes v_{0}'.
\end{equation}

\begin{table*}
\begin{ruledtabular}
  \caption{  
    \label{tab: a1 su4} Multiplication table for the elements of the algebra associated to $A_1$ of $SU(4)$. For simplicity, it is assumed that before evaluating these expressions, one must take the $\pmod{4}$ of each index.}
    \begin{tabular}{ccccc}
  
        $\cdot$ & $z_i \otimes z_j$ & $x^{(1)}_i \otimes x^{(1)}_j$ & $x^{(2)}_i \otimes x^{(2)}_j$ & $y^{(1)}_i \otimes y^{(1)}_j$  \\ \hline
       $z_l \otimes z_m$  & $\delta_{li} \delta_{mj} (z_l \otimes z_m)$ & $\delta_{li} \delta_{mj}(x^{(1)}_i \otimes x^{(1)}_j)$ & $\delta_{li} \delta_{mj}(x^{(2)}_i \otimes x^{(2)}_j)$ & $\delta_{(l+3)i} \delta_{(m+3)j} (y^{(1)}_i \otimes y^{(1)}_j)$  \\
        
       $x^{(1)}_l \otimes x^{(1)}_m$ & $\delta_{(l+1)i} \delta_{(m+1)j}(x^{(1)}_l \otimes x^{(1)}_m)$ & $\delta_{(l+1) i} \delta_{(m+1)j}(x^{(2)}_{l}\otimes x^{(2)}_{m})$ & $\delta_{(l+1)i} \delta_{(m+1)j} (y^{(1)}_{l+3} \otimes y^{(1)}_{m+3})$ & $\delta_{li} \delta_{mj} (z_l \otimes z_m)$\\ 
     
       $x^{(2)}_l \otimes x^{(2)}_m$ & $\delta_{(l+2)i} \delta_{(m+2)j} (x^{(2)}_l \otimes x^{(2)}_m)$& $\delta_{(l+2)i} \delta_{(m+2)j} (y^{(1)}_{l+3} \otimes y^{(1)}_{m+3})$ & $\delta_{(l+2)i} \delta_{(m+2)j} (z_l \otimes z_m)$  & $\delta_{(l+1)i}\delta_{(m+1)j}(x^{(1)}_{l} \otimes x^{(1)}_{m})$ \\ 
       
       $y^{(1)}_l \otimes y^{(1)}_m$ & $\delta_{li} \delta_{mj} (y^{(1)}_l \otimes y^{(1)}_m)$ & $\delta_{li} \delta_{mj} (z_{l+1} \otimes z_{m+1})$ & $\delta_{li} \delta_{mj} (x^{(1)}_{l+1} \otimes x^{(1)}_{m+1})$ & $\delta_{(l+3)i} \delta_{(m+3)j} (x^{(2)}_i \otimes x^{(2)}_j)$ \\
    \end{tabular}
    \end{ruledtabular}
\end{table*}

Multiplication by a vertex is trivial 
\begin{equation}
    (v_{0}\otimes v'_{0})\cdot (v_{1}\otimes v_{1}')=\delta_{v_{0}v_{1}}\delta_{v_{0}'v_{1}'}(v_{0}\otimes v_{0}'),
\end{equation}
\begin{equation}
    (v_0 \otimes v_0') \cdot (v_1 v_2 \otimes v_1' v_2') = \delta _{v_0 v_1} \delta_{v_0' v_1'} (v_1 v_2 \otimes v_1' v_2'),
\end{equation}
\begin{equation}
     (v_0 v_1 \otimes v_0' v_1')\cdot (v_2 \otimes v_2') = \delta _{v_1 v_2} \delta_{v_1' v_2'} (v_0 v_1 \otimes v_0' v_1'),
\end{equation}
The rest of products not listed here are zero or are products of non-graded endomorphisms and therefore, not relevant for this work.

\section{About higher level $A_l$ graphs and final discussion}
\label{section: higherlevelA}
In this work, the $SU(N)$ $A_1$ associated quantum groupoid has been successfully constructed by finding the decomposition of each $\mathcal{P}_{n}$ for any $N$. Given that the operators defined act on path sections with $2$ steps or less (which also appear in the rest of the $A_{l}$ $SU(3)$ graphs), and given that almost all the proofs in this work can be done without supposing a particular graph, it seems natural to extend this construction to other levels. 

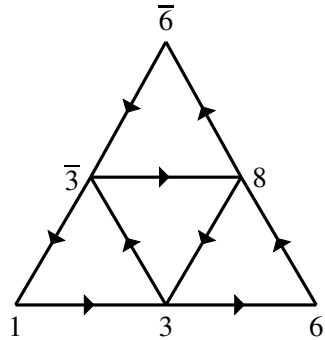
\begin{figure}[H]
\centering
\begin{tikzpicture}
\begin{scope}[very thick, every node/.style={sloped,allow upside down}]
  \draw (0,0) node[below] {1}-- node {\midarrow} (2,0);
  \draw (2,0) node[below] {3}-- node {\midarrow} (4,0) ;
  \draw (4,0) node[below] {6} -- node {\midarrow} (3,1.7);
  \draw (3,1.7) node[right] {8} -- node {\midarrow} (2,3.5);
  \draw (2,3.5) node[above] {$\overline{6}$} -- node {\midarrow} (1,1.7);
  \draw (1, 1.7) node[left] {$\overline{3}$} -- node {\midarrow} (0,0);
  \draw (2,0) -- node {\midarrow} (1,1.7);
  \draw (3, 1.7) -- node{\midarrow} (2,0) ;
  \draw (1, 1.7) -- node{\midarrow} (3,1.7);
\end{scope}
\end{tikzpicture}
\caption{The $SU(3)$ $A_2$ graph generated by the fundamental representation $(1,0)$ with the vertices labeled by quantum dimensions.}
\label{A_2 graph}
\end{figure}

However, when working with $A_l$ graphs with $l>1$, some new $2$-step paths arise, which are neither $2$-step loops nor $2$ sides of a triangle. Some of them are straight lines, for example, in the $SU(3)$ $A_{2}$  graph, there are paths like $(136)$. There are also paths which are two sides of a parallelogram, as $(138)$ and $(1 \overline{3}8)$ (see figure \ref{A_2 graph}). The set of creation operators previously defined do not account for paths like $(138)$, especially because it is unclear whether they are essential or not.

Since the shortest paths connecting $1$ and $8$ are $(138)$ and $(1\bar{3}8)$, these are essential given our previous definitions. Nonetheless, if this is the case, there would be essential paths with $3$ steps such as $(1\bar{3}86)$. In $A_{2}$, the maximum number of steps of an essential path is $2$, thus, there is a contradiction. In addition, the number of essential paths do not match with the one predicted by the references.

Moreover, after some trial an error, both the decomposition of the subspaces of paths $\mathcal{P}_{n}$, and the realization of the Temperley-Lieb algebra seem to be unreachable.

So, it seems that a piece of information is missing for the higher levels. Given that each type of $2$-step path has its own pair of associated operators, it would be natural to define some operators which deal with these problematic paths. Furthermore, this situation is similar to the one presented in quantum mechanics, when our set of operators is not a CSCO and there are degenerate states. In other words, there should be only one essential path of length $(1,1)$ which connects $1$ and $8$ and the set of operators we need to define should allow us to identify it. However, each operator seems to be related to a combination of two generators, and we do not have more generators on $SU(3).$ 

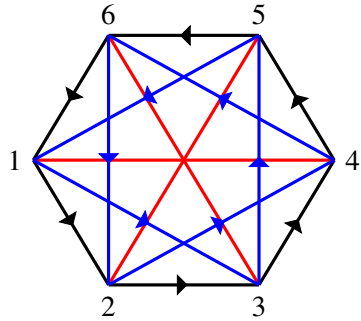
\begin{figure}[H]
\centering
\begin{tikzpicture}
\begin{scope}[very thick, every node/.style={sloped,allow upside down}]
  \draw (-2,0) node[left] {1}-- node {\midarrow} (-1,-1.66);
  \draw (-1,-1.66) node[below] {2} -- node {\midarrow} (1,-1.66);
  \draw (1,-1.66) node[below]{3} -- node {\midarrow} (2,0);
  \draw (2,0) node[right] {4} -- node {\midarrow} (1,1.66);
  \draw (1,1.66) node[above] {5} -- node {\midarrow} (-1,1.66);
  \draw (-1,1.66) node[above] {6} -- node {\midarrow} (-2,0);
   \draw[red] (-2,0) --  (2,0);
   \draw[red] (-1,-1.66) --  (1,1.66);
   \draw[red] (1,-1.66) --  (-1,1.66);
  \draw[blue] (-2,0) -- node{\midarrow} (1,-1.66);
  \draw[blue] (1,-1.66) -- node{\midarrow} (1,1.66);
  \draw[blue] (1,1.66) -- node{\midarrow} (-2,0);
  \draw[blue] (-1,1.66) -- node {\midarrow} (-1,-1.66);
  \draw[blue] (-1,-1.66) -- node {\midarrow} (2,0);
  \draw[blue] (2,0) -- node {\midarrow} (-1,1.66);
 
\end{scope}
\end{tikzpicture}
\caption{\label{hexagono}The $A_1$ $SU(6)$ graph. $SU(6)$ has $5$ fundamental generators, one corresponding to the black oriented segments (and its conjugate), one corresponding to the blue oriented segments (and its conjugate) and one corresponding to the red segments (which is its own conjugate) \cite{Robertpage}. }
\end{figure}

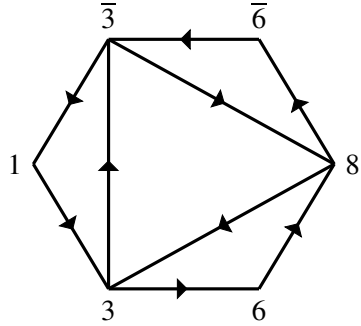
\begin{figure}[H]
\centering
\begin{tikzpicture}
\begin{scope}[very thick, every node/.style={sloped,allow upside down}]
  \draw (-2,0) node[left] {1}-- node {\midarrow} (-1,-1.66);
  \draw (-1,-1.66) node[below] {3} -- node {\midarrow} (1,-1.66);
  \draw (1,-1.66) node[below]{6} -- node {\midarrow} (2,0);
  \draw (2,0) node[right] {8} -- node {\midarrow} (1,1.66);
  \draw (1,1.66) node[above] {$\overline{6}$} -- node {\midarrow} (-1,1.66);
  \draw (-1,1.66) node[above] {$\overline{3}$} -- node {\midarrow} (-2,0);
  \draw (-1,-1.66) -- node {\midarrow} (-1,1.66);
  \draw (2,0) -- node {\midarrow} (-1,-1.66);
  \draw (-1,1.66) -- node {\midarrow} (2,0);
\end{scope}
\end{tikzpicture}
\caption{\label{hexagonomodificado }This is the graph obtained after truncating figure (\ref{hexagono}). The process of truncation required to remove some edges and to invert some of the remaining ones. The number of different generators is also reduced and we have relabeled the vertices to match the $A_1$ $SU(3)$ graph.}
\end{figure}

On the other hand, notice that the $SU(3)$ $A_2$ graph can be realized as a truncation of the $SU(6)$ $A_{1}$ graph (see the changes from figure \ref{hexagono} to figure \ref{hexagonomodificado }). 

This lead us to think that the paths like $(138)$ in the $SU(3)$ $A_{2}$ graph are actually produced by applying the triangle operators of $SU(6)$ to edges like $(18)$ which is not an edge belonging to the $SU(3)$ $A_{2}$ graph. If that is the case, we do not need more operators because  we can obtain a natural decomposition of each $\mathcal{P}_{n}$ for the $SU(3)$ $A_{2}$ graph by taking the decomposition of each $\mathcal{P}_{n}$ of the $SU(6)$ $A_{1}$ graph and neglecting the paths which are not part of the $SU(3)$ $A_{2}$ graph.

However, this decomposition seems to hide the essential paths of the $SU(3)$ $A_{2}$ graph, which make the construction of the $C^\star$ bialgebra harder. But, as mentioned before, it suggests us that the triangle operators of $SU(6)$ are actually the responsible for the paths like $(138)$ and $(1\bar{3}8)$ in the $SU(3)$ $A_2$ graph. Also, it suggests us that the decomposition of each $\mathcal{P}_{n}$ is possible and therefore, that the $C^{\star}$ bialgebra associated to $SU(3)$ $A_{2}$ graph can actually be constructed. 

Moreover, we can manipulate that natural decomposition to get another one in which the essential paths are not hidden anymore. We just need to realize how to define the creation operators associated to the $SU(3)$ $A_{2}$ graph in terms of the creation operators acting over paths on the $SU(6)$ $A_{1}$ graph, and a rule to distinguish essential paths from non-essential paths. Some progress has already been made towards this direction, and the generalization to all the $SU(N)$ $A_{l}$ graphs is on its way.


 Once we find the decomposition for any level, we will be able to reproduce the construction of the $C^\star$ bialgebra in general. This is not an easy task, but there are good chances to finally find the generalization for the $A_l$ $SU(N)$ associated quantum groupoids.


\bibliography{bibliography-paths-su3}

\end{document}